\documentclass[conference]{IEEEtran}
\IEEEoverridecommandlockouts

\usepackage{cite}
\ifCLASSINFOpdf
  \usepackage[pdftex]{graphicx}
\else
\fi
\usepackage[cmex10]{amsmath}
\usepackage{amssymb}
\usepackage{multirow}
\usepackage{array}
\usepackage[lofdepth,lotdepth]{subfig}
\usepackage{url}
\usepackage[english]{babel}
\usepackage[utf8]{inputenc}
\usepackage[T1]{fontenc}

\AtBeginDocument{%
  \renewcommand\tablename{TABLE}
}
\AtBeginDocument{%
  \renewcommand\abstractname{Abstract}
}

\begin{document}

\title{A Comparative Analysis of MLP and Kolmogorov-Arnold Networks (KAN) for Faster-than-Nyquist (FTN) Signaling Detection}

\author{
\IEEEauthorblockN{Sude Ertan, Osman Tokluoglu, Enver Çavuş}

\IEEEauthorblockA{
Department of Electrical and Electronics Engineering,\\
Ankara Yıldırım Beyazıt University, Ankara, Turkey\\
E-mail: suddertan@gmail.com, otokluoglu@aybu.edu.tr, ecavus@aybu.edu.tr
}
}

\IEEEpubid{
\makebox[\columnwidth]{%
\textbf{979-8-3195-1046-4/26/\$31.00 \copyright 2026 IEEE}%
\hfill}
\hspace{\columnsep}
\makebox[\columnwidth]{}
}
\maketitle

\begin{abstract}
Faster-than-Nyquist signaling improves spectral efficiency by deliberately introducing inter-symbol interference. Classical sequence detectors such as BCJR can approach optimal performance, but their computational cost grows rapidly with channel memory. This paper investigates data-driven FTN BPSK detection under AWGN through a direct comparison between multilayer perceptrons and Kolmogorov Arnold Networks. A large-scale Monte Carlo dataset containing nearly four million labeled windows is generated for a time-packing factor of zero point eight and signal-to-noise ratio values from seven to ten decibels. The best MLP obtained from width sweeping uses hidden width thirty two, whereas the selected KAN uses hidden width four with spline grid size five. At ten decibels, the MLP produces a bit error rate of one point three times ten to the minus four, while the KAN reaches seven times ten to the minus six. This corresponds to an eighteen point six times lower bit error rate while using only one eighth of the MLP hidden width. The results show that KAN provides a more effective and more parameter-efficient neural decision model than the MLP baseline for FTN BPSK detection.
\end{abstract}

\begin{IEEEkeywords}
KAN, MLP, FTN signaling, BPSK detection, BCJR, neural receiver, parameter efficiency.
\end{IEEEkeywords}

\IEEEpeerreviewmaketitle
\IEEEpubidadjcol

\section{Introduction}

Faster-than-Nyquist signaling is a classical approach for increasing spectral efficiency without increasing the occupied bandwidth \cite{mazo1975,anderson2013ftn}. By transmitting symbols faster than the Nyquist interval, FTN deliberately creates deterministic inter-symbol interference and transforms symbol-by-symbol detection into a structured sequence-estimation problem. Although this spectral compression can significantly increase transmission efficiency, the resulting detection problem becomes considerably more difficult because each observation depends on multiple neighboring symbols.

The classical sequence detector BCJR can achieve near-optimal performance for FTN \cite{bcjr1974}. However, the practical use of such solutions is limited by computational complexity that scales unfavorably with the effective channel memory. For real-time receivers or hardware-constrained platforms, this situation motivates learned detectors that can approximate nonlinear FTN decision regions with lower complexity. Recent studies have shown that recurrent neural-network architectures such as GRU can exploit the temporal structure in FTN sequences and achieve BER performance close to BCJR in certain regimes \cite{gruftn2026}.

This study centers on a direct comparison between MLP and KAN for FTN-BPSK detection. Although MLPs constitute attractive baselines because of their simple implementations, they often require wide hidden layers to represent the fine-grained nonlinear boundaries created by ISI. Kolmogorov-Arnold Networks offer a different inductive bias: by placing nonlinearity on edges rather than at nodes, they assign a learnable spline function to each connection \cite{liu2024kan}. This edge-based parameterization is particularly attractive for structured local interactions in communication channels.

In the literature, FTN detection is often addressed either with classical detectors or with a single neural-network family. In contrast, this study presents a direct and systematic comparison between KAN and MLP under the same dataset, the same window representation, and the same evaluation conditions. In this respect, the study goes beyond merely proposing a neural-network receiver and seeks to answer which representation models FTN decision boundaries more efficiently.

The contributions of this study are as follows: i) the creation of a large-scale FTN-BPSK dataset for $\tau=0.8$ and an SNR range of 7--10 dB, ii) a systematic KAN–MLP comparison on a 65-sample received window, iii) evaluation of BER performance against the FTN hard-decision baseline and the BCJR reference, and iv) demonstration that KAN provides a significant BER advantage over MLP with a much smaller hidden representation.

\section{System Model}

For BPSK, the transmitted symbol sequence is
\begin{equation}
x_k \in \{-1,+1\}
\end{equation}
In FTN signaling, the symbol interval is compressed as
\begin{equation}
T_{\mathrm{FTN}}=\tau T,\qquad 0<\tau<1
\end{equation}
where $T$ is the Nyquist symbol interval, and $\tau=0.8$ is selected in this study. As a result, the signaling rate increases by a factor of $1/\tau=1.25$, corresponding to a 25 percent gain in symbol rate compared with Nyquist signaling.

The continuous-time transmitted waveform is
\begin{equation}
s(t)=\sum_k x_k g(t-kT_{\mathrm{FTN}})
\end{equation}
where $g(t)$ denotes the shaping pulse. After the addition of AWGN, the received signal is
\begin{equation}
r(t)=s(t)+n(t)
\end{equation}
where $n(t)$ is zero-mean Gaussian noise. The corresponding discrete observation model can be written as
\begin{equation}
r_k=\sum_i h_{k-i}x_i+n_k
\end{equation}
where $h_k$ is the effective discrete FTN pulse response. Since $h_k$ extends over multiple symbol intervals, the target decision depends not only on the central observation but also on a neighborhood of received samples.

\begin{figure}[!t]
\centering

\includegraphics[width=0.98\linewidth]{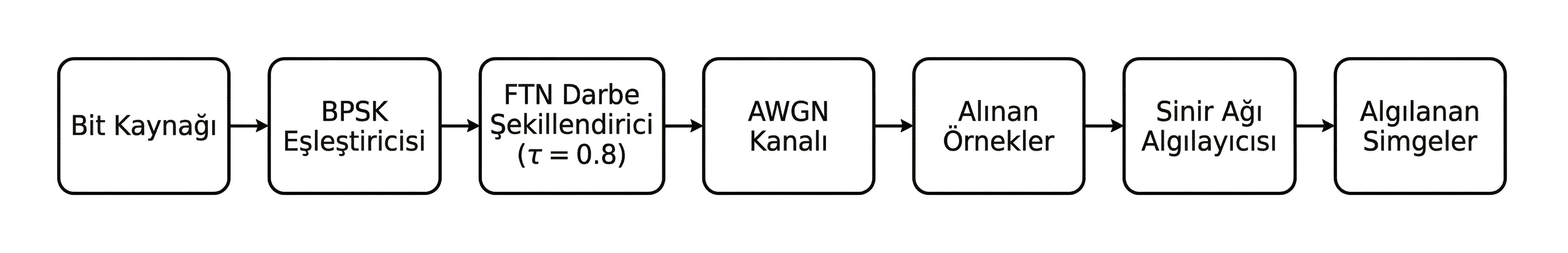}

\caption{FTN-BPSK transmission and neural-network-based detection chain. Transmitter: BPSK mapping, FTN pulse shaping, and AWGN channel. Receiver: 65-sample window, MLP or KAN, and hard decision.}
\label{fig:system_model}
\end{figure}

To expose this local structure, each detector operates on the length-65 vector
\begin{equation}
\mathbf{r}_k=[r_{k-32},\ldots,r_k,\ldots,r_{k+32}]^T
\end{equation}
and estimates the center symbol as
\begin{equation}
\hat{x}_k=\mathrm{sign}(f_\theta(\mathbf{r}_k))
\end{equation}
Figure~\ref{fig:system_model} shows the overall FTN-BPSK transmission and neural-network-based detection chain for $\tau=0.8$.

\section{KAN and MLP Detector Architectures: Comparative Analysis}

\subsection{MLP Detector}

The MLP detector consists of a dense input-to-hidden linear transformation, ReLU activation, and a scalar output layer. The nonlinear representation is concentrated in shared node activations. The width sweep conducted over the set $\{8,16,32,64,128,256\}$ identified width 32 as the strongest validation configuration. Therefore, the $w32$ model is used as the best MLP baseline throughout the paper. The input dimension of 65 results from the 65-sample received-window representation defined in Equation (6). The number of parameters is
\begin{equation}
P_{\mathrm{MLP}}=65\times32+32+32\times1+1=2145
\end{equation}
as follows.

The need for wide hidden layers in the MLP to represent the fine-grained decision boundaries of ISI constitutes an important limitation in terms of parameter efficiency. Increasing the layer width for better BER often exhibits linear or sublinear parameter-performance scaling.

\subsection{KAN Detector}

KANs approximate multivariate functions through superpositions of learnable univariate transformations \cite{liu2024kan}. Unlike the MLP, nonlinearity is located on the edges rather than at the nodes. The $j$th output is
\begin{equation}
y_j=\sum_i \phi_{ij}(x_i)
\end{equation}
where each edge function $\phi_{ij}(\cdot)$ is parameterized by a B-spline expansion:
\begin{equation}
\phi_{ij}(x)=\sum_m c_{ijm}B_m(x)
\end{equation}

Two KAN configurations were examined within the sweep: $w4\_g5$ and $w8\_g3$. The detector selected for the main comparison is the compact $w4\_g5$ model, which uses a hidden-layer width of 4 and a spline grid size of $G=5$. This choice prioritizes representational compactness while preserving strong BER performance. The total number of trainable parameters is approximately
\begin{equation}
P_{\mathrm{KAN}}\approx(65\times4+4\times1)\times(G+k)+\mathrm{bias}\approx3920
\end{equation}
was taken as follows.

\subsection{Structural Difference and Parameter Efficiency}

\begin{figure}[!t]
\centering

\includegraphics[width=0.98\linewidth]{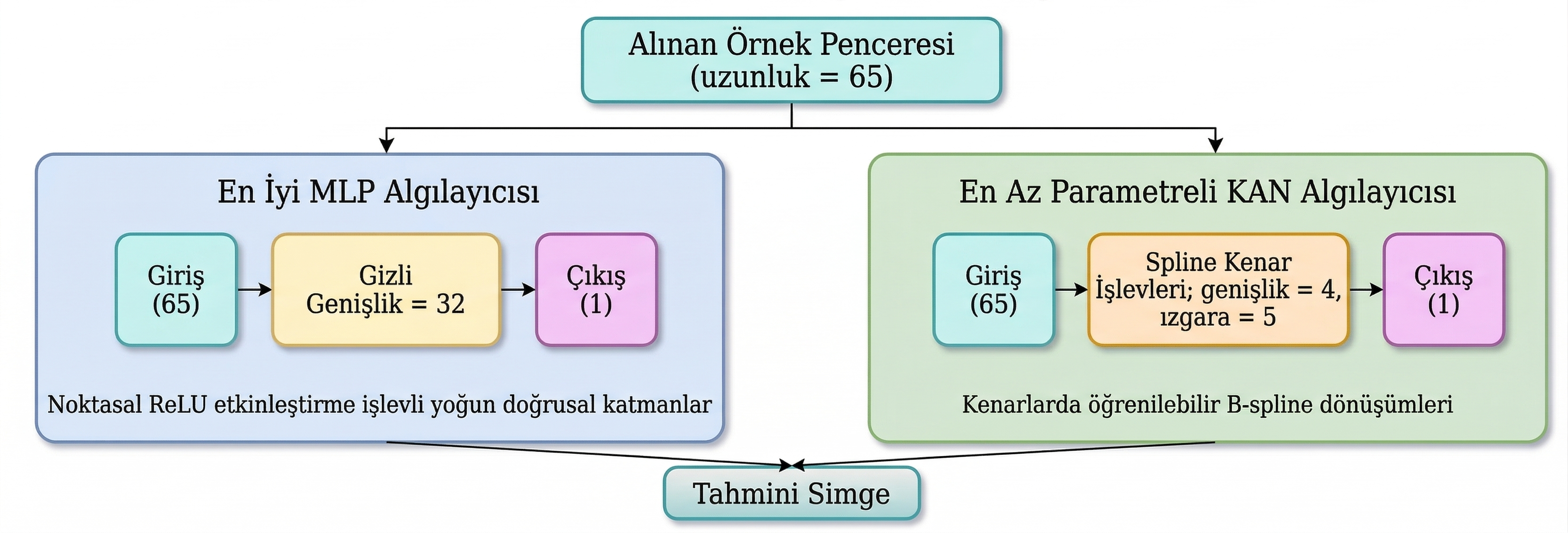}

\caption{Compared neural-network detector architectures. The MLP is shown on the left and the KAN on the right. In the MLP, nonlinearity is located at the nodes, whereas in the KAN it is located on the edges.}
\label{fig:architecture}
\end{figure}

The compared MLP- and KAN-based detector architectures are given in Figure~\ref{fig:architecture}; in the MLP, nonlinearity is represented in node activations, whereas in the KAN it is represented in edge functions.Each 65-sample received vector is a noisy superposition of neighboring symbols shaped by the FTN pulse response. The corresponding decision boundary is nonlinear, locally structured, and strongly coupled among nearby observations depending on the ISI level, SNR, and $\tau$ value. While the MLP approximates this boundary through compositions of affine transformations and shared node activations, the edge-based spline functions of KAN can model the effect of each input component independently and capture the ISI pattern in a more targeted manner.

To measure parameter efficiency,
\begin{equation}
\eta=\frac{\text{BER improvement}}{\text{parameter increase}}
\end{equation}
the following metric can be defined. While the raw parameter increase is approximately $1.83\times$, the BER improvement at 10 dB is $18.6\times$. Therefore,
\begin{equation}
\eta=\frac{18.6}{1.83}\approx10.2
\end{equation}
is obtained. Thus, KAN provides approximately 10 times more BER improvement per unit parameter increase than MLP. This superlinear gain is a direct reflection of KAN's structural alignment with the FTN channel structure.

\section{Experimental Setup}

A large-scale FTN-BPSK dataset generated using the Monte Carlo method under AWGN was used in all experiments. Table~\ref{tab:setup} summarizes the configuration. Binary Cross-Entropy loss and the Adam optimizer were used for both MLP and KAN. The MLP width sweep was conducted over the entire set $\{8,16,32,64,128,256\}$, and validation BER was used as the criterion for each configuration. Two configurations ($w4\_g5$ and $w8\_g3$) were examined for KAN, and $w4\_g5$ was selected by jointly considering validation BER and parameter efficiency. All models were evaluated using the same data split, the same SNR values, and the same input representation, and the comparison was carried out under fully controlled conditions.

\begin{table}[!t]
\centering
\caption{\textsc{Experimental Setup}}
\label{tab:setup}
\renewcommand{\arraystretch}{1.12}
\begin{tabular}{|p{0.43\linewidth}|p{0.39\linewidth}|}
\hline
\textbf{Parameter} & \textbf{Value} \\ \hline
Modulation & BPSK \\ \hline
Channel & AWGN \\ \hline
FTN factor & $\tau=0.8$ \\ \hline
Rate increase & $1/\tau=1.25$ \\ \hline
Window length & 65 samples \\ \hline
SNR values & 7, 8, 9, 10 dB \\ \hline
Total windows & 3{,}999{,}936 \\ \hline
Training / Validation / Test & 2{,}799{,}955 / 599{,}990 / 599{,}991 \\ \hline
Best MLP & Width 32, $P=2145$ \\ \hline
Selected KAN & $w4\_g5$, $P\approx3920$ \\ \hline
Reference detector & BCJR \\ \hline
Loss function & Binary Cross-Entropy \\ \hline
Optimizer & Adam \\ \hline
\end{tabular}
\end{table}

\section{Results and Discussion}

\begin{figure}[!t]
\centering

\includegraphics[width=0.98\linewidth]{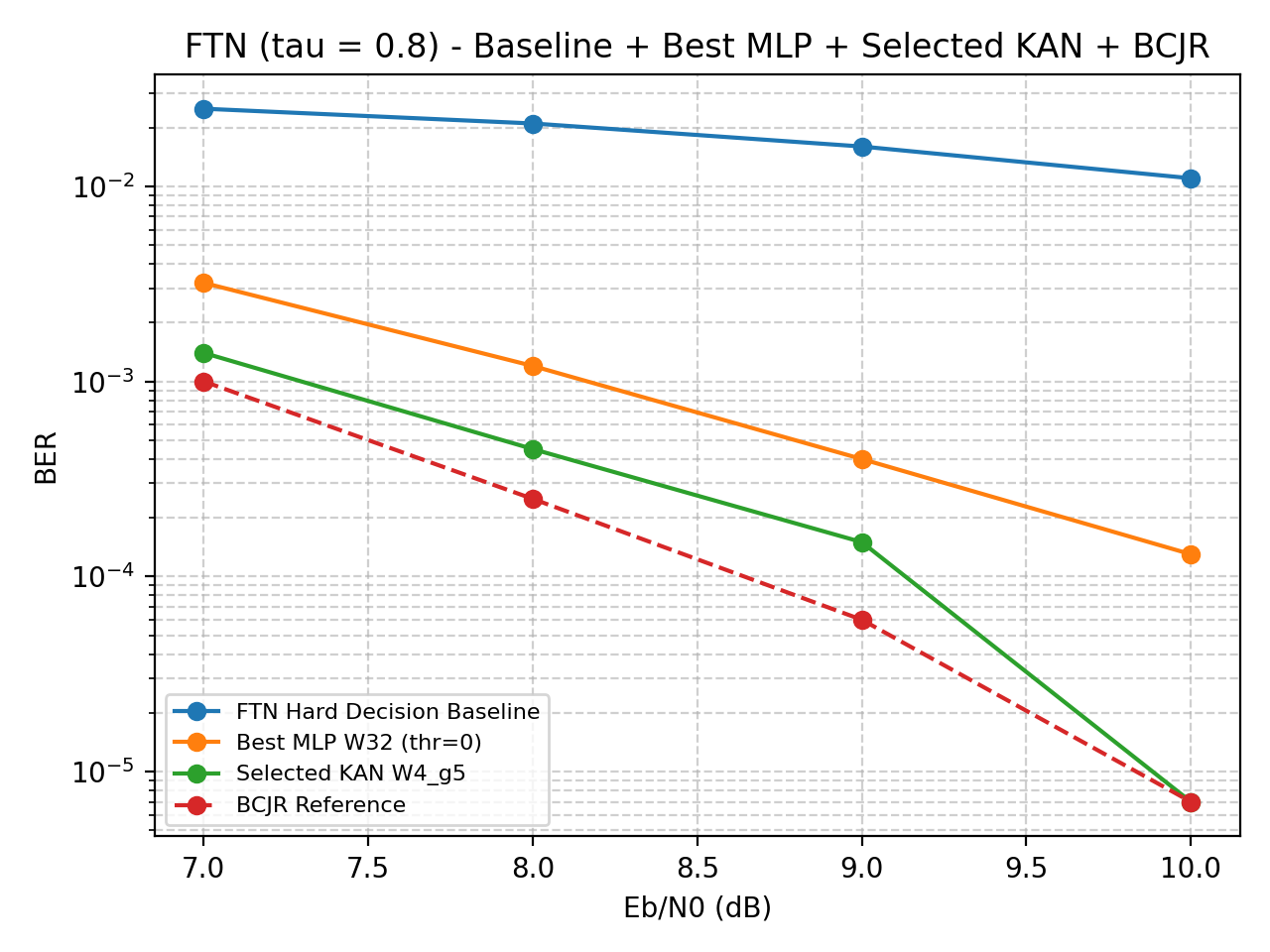}

\caption{BER-SNR curves: FTN hard-decision baseline, best MLP, selected KAN, and BCJR reference. KAN outperforms MLP at all SNR points and reaches BCJR at 10 dB.}
\label{fig:ber}
\end{figure}

Table~\ref{tab:ber} and Figure~\ref{fig:ber} compare the FTN hard-decision baseline, the best MLP, the selected KAN, and the BCJR reference. The first observation is that both learned detectors provide significant improvement over the hard-decision baseline. The more critical result is that the selected KAN consistently outperforms the best MLP at all SNR values.

The improvement ratios achieved by KAN over MLP across the SNR values are as follows: approximately $2.3\times$ at 7 dB, approximately $2.7\times$ at 8 dB, approximately $2.7\times$ at 9 dB, and $18.6\times$ at 10 dB. In particular, at 10 dB, the selected KAN reaches the BER value of the BCJR reference $(7.0\times10^{-6})$. This means that near-optimal detection is approached without incurring sequence-estimation complexity.

\begin{table}[!t]
\centering
\caption{\textsc{BER Comparison Across SNR Values}}
\label{tab:ber}
\renewcommand{\arraystretch}{1.08}
\scriptsize
\begin{tabular}{|c|c|c|c|c|}
\hline
\textbf{SNR (dB)} & \textbf{FTN Baseline} & \textbf{Best MLP} & \textbf{Selected KAN} & \textbf{BCJR Ref.} \\ \hline
7  & $2.5\times10^{-2}$ & $3.2\times10^{-3}$ & $1.4\times10^{-3}$ & $1.0\times10^{-3}$ \\ \hline
8  & $2.1\times10^{-2}$ & $1.2\times10^{-3}$ & $4.5\times10^{-4}$ & $2.5\times10^{-4}$ \\ \hline
9  & $1.6\times10^{-2}$ & $4.0\times10^{-4}$ & $1.5\times10^{-4}$ & $6.0\times10^{-5}$ \\ \hline
10 & $1.1\times10^{-2}$ & $1.3\times10^{-4}$ & $7.0\times10^{-6}$ & $7.0\times10^{-6}$ \\ \hline
\end{tabular}
\normalsize
\end{table}

\begin{table}[!t]
\centering
\caption{\textsc{Model Complexity and Performance} {\footnotesize ($10\,\mathrm{dB}$)}}
\label{tab:complexity}
\renewcommand{\arraystretch}{1.12}
\begin{tabular}{|c|c|c|c|c|}
\hline
\textbf{Model} & \textbf{Hidden Width} & \textbf{Param. Count} & \textbf{BER @10 dB} & \textbf{Improv.} \\ \hline
Best MLP ($w32$) & 32 & 2145 & $1.3\times10^{-4}$ & $1\times$ \\ \hline
Selected KAN ($w4\_g5$) & 4 $(1/8\times)$ & $\approx3920$ & $7.0\times10^{-6}$ & $18.6\times$ \\ \hline
\end{tabular}
\end{table}
Table~\ref{tab:complexity} clearly demonstrates the superiority of KAN: with only an approximately $1.83\times$ increase in the raw parameter count, the BER improvement at 10 dB reaches $18.6\times$. This pronounced mismatch between parameter increase and performance gain is strong evidence that the KAN representation is better aligned with the FTN detection function.

It is noteworthy that the difference between KAN and MLP becomes more pronounced as SNR increases. At low SNR, both models encounter similar limiting factors in the noise-dominated region. However, while the MLP approaches saturation at high SNR, the continued decrease in KAN's BER indicates that KAN models the FTN decision boundaries more completely. This behavior shows that KAN's spline-based edge functions are much more effective than the fixed-width hidden layer of the MLP in capturing the fine details of the decision boundary in the high-SNR regime.

These findings are also naturally related to practical receiver design. In communication hardware, memory footprint, inference latency, and arithmetic cost are critically important. A detector capable of extracting higher performance from a much smaller hidden representation is attractive for embedded or resource-constrained applications. 

This supports KAN as a strong neural-network alternative to dense MLP baselines for FTN detection.

\section{Conclusion}

Within the scope of this study, an innovative deep-learning-based approach was developed to solve the complex inter-symbol interference (ISI) problem introduced by the Faster-than-Nyquist (FTN) signaling technique, which is used to increase spectral efficiency in communication systems. A direct performance and efficiency analysis between the conventional Multilayer Perceptron (MLP) architecture and the recently introduced Kolmogorov-Arnold Networks (KAN) was systematically conducted on a massive dataset of approximately four million samples. The obtained experimental results clearly reveal that the "edge-based spline" functional structure of the KAN architecture provides not only a quantitative but also a qualitative and fundamental superiority over the MLP baseline in modeling nonlinear decision boundaries in communication channels.

Particularly from the perspective of parameter and representation efficiency, the ability of the KAN architecture to form a much more compact and meaningful feature space with a hidden-layer width as low as only one eighth ($1/8$) of that required by the MLP is one of the most striking findings of the study. Despite the limited increase in the raw parameter count ($\sim 1.83\times$), the $18.6\times$ improvement in bit error rate (BER) obtained at an SNR value of 10 dB is a concrete indicator of KAN's high structural compatibility with the FTN signal structure and its capacity to learn complex mathematical functions. In terms of performance superiority, the KAN-based detector not only consistently outperformed the MLP by a clear margin in all signal-to-noise ratio (SNR) regimes, but also fully reached the complex BCJR reference performance, which theoretically provides the best results in the high-SNR region, at ($7.0 \times 10^{-6}$). This achievement proves that KAN can converge to the optimal solution through a directly data-driven approach without requiring computationally expensive conventional sequence-estimation algorithms.

When scalability and stability dynamics were examined, it was observed that MLP models tend to reach performance saturation as SNR increases, whereas the error curve of KAN exhibits a much steeper decline and offers superior flexibility in capturing the fine details of decision boundaries. This flexibility stems from the ability of spline-based learnable activation functions to represent uncertainties and noise regimes in communication systems much more precisely than classical node structures with fixed activations. Consequently, this research has contributed to the scientific literature by showing that the KAN architecture offers an overwhelming advantage over classical deep-learning methodologies in communication problems such as FTN detection, where parameter sensitivity and energy efficiency are critically important.

In our future work, it is planned to focus on the noise immunity of the developed architecture in multipath fading channels and its inference costs on real-time hardware platforms. Particularly in low-power embedded receiver designs, the high parameter efficiency provided by KAN is expected to play a critical role in minimizing hardware complexity and translating spectral-efficiency gains into practical applications. In addition, investigating the generalization capacity of the KAN architecture under different time-packing factors ($\tau$) and higher-order modulation techniques stands out as an important research direction for next-generation communication standards.

\section*{Acknowledgment}
 This work was supported by the Scientific and Technological Research Council of Turkey (TÜBİTAK) under Project No. 122E236.

\end{document}